\documentclass[aps,pra,onecolumn,preprint,preprintnumbers,groupaddress,amsmath,amssymb,usenatbib,
nofootinbib,showkeys,showpacs,altaffilletter,11pt]{revtex4}
\usepackage{graphicx}
\usepackage{dcolumn}
\usepackage{bm}
\usepackage{latexsym}
\usepackage{amsfonts}
\usepackage{amssymb}
\usepackage{amsmath}
\usepackage{subfigure}
\usepackage{mathrsfs}
\usepackage{tabu}
\usepackage{xcolor}
\usepackage{soul}

\begin{document}

\title{Thermodynamics of the viscous $f(T, B)$  gravity in the new agegraphic dark energy model}

\author{A. Pourbagher}
\email[]{st.a.pourbagher@iauamol.ac.ir}
\author{Alireza Amani}
\email[Corresponding author: ]{a.r.amani@iauamol.ac.ir}
\affiliation{Department of Physics, Ayatollah Amoli Branch, Islamic Azad University, Amol, Iran}

\date{\today}

\begin{abstract}
       In this paper, we first obtain the energy density by the approach of the new agegraphic dark energy model, and then the $f(T,B)$ gravity model is studied as an alternative to the dark energy in a viscous fluid by flat-FRW background, in which $T$ and $B$ are torsion scalar and boundary term. The Friedmann equations will obtain in the framework of modified teleparallel gravity by tetrad components. We consider that the universe dominates with components such as matter and dark energy by an interacting model. The Hubble parameter is parameterized by the power-law for the scale factor, and then we fit the corresponding Hubble parameter with observational data constraints. The variation of the equation of state (EoS) for dark energy is plotted as a function of the redshift parameter, and the accelerated expansion of the universe is explored.  In what follows, the stability of the model is also studied on the base of the sound speed parameter. Finally, the generalized second law of thermodynamics is investigated by entropies of inside and on the boundary of the apparent horizon in thermodynamics equilibrium.
\end{abstract}
\maketitle

\section{Introduction}\label{I}

 Research conducted over the past two decades cleared that the universe is undergoing accelerated expansion in which the speed of the expansion of the universe is continually increasing with time.  This was first observed in type Ia supernova data and then reconfirmed using the cosmic microwave background and the large scale structure \cite{Riess_1998, Perlmutter_1999, Bennett_2003, Tegmark_2004}. The reason for this expansion lies in the definition of a kind of magic energy called dark energy which is three-quarters of the total energy of the universe. This expansion will occur if the universe contains negative-pressure energy sources until dark energy can be described. To prove this subject, efforts have been made to understand the cause of acceleration and many theoretical models have presented to describe it. Some of these models are the cosmological constant, the scalar fields (quintessence, phantom, tachyon, and so on), the modified gravity models, the bouncing model, and the braneworld models \cite{Weinberg-1989, Caldwell-2002, Amani-2011, Sadeghi1-2009, Setare-2009, Setare1-2009, Battye-2016, Li-2012, Pourhassan-2014, JSadeghi-2015, Amani1-2015, Iorio-2016, Faraoni-2016, Khurshudyan1-2014, Sadeghi1-2016, Wei-2009, Amani-2015, Nojiri_2007, Li-2004, Campo-2011, Hu-2015, Fayaz-2015, Saadat1-2013, Amani-2013, Amani-2014, Naji-2014, Morais-2017, Zhang1-2017, KhurshudyanJ2-2014, SadeghiKhurshudyanJ1-2014, Sadeghi-2010, Sadeghi-2009, Amani-2016, Singh-2016, Sahni1-2003, Setare-2008, Brito-2015, Setaremr-2008, Amanifarahani-2012, Amanifarahani1-2012, Amanipourhassan-2012,  Bhoyar-2017, Chirde-2018, Singh-2018, Nagpal-2018}.

Also, this subject as a fundamental theory in string theory and quantum gravity is appeared, though the theory of quantum gravity has not yet been completed, there are many attempts to investigate the nature of dark energy based on some principles of quantum gravity recently. The holography and agegraphic are two examples of such models that originated from some features of quantum gravity theory \cite{Amani1-2011, Wei-2008}.

In this paper, we investigate the agegraphic dark energy with uncertainty in quantum mechanics by the gravitational effect which derived from general relativity. Here, it is assumed that the agegraphic dark energy model arises from space-time and the fluctuations of the matter field in the universe. Thus, in this model, instead of the horizon distance, which is the problem of causality in the holographic dark energy, the age of the universe is chosen as the length of the universe \cite{Wei-2008}. This model has also been able to adapt to astronomical observations \cite{Karolyhazy-1966, Karolyhazy-1982, Karolyhazy-1986, Maziashvili-2007, Maziashvili1-2007, Wei-2007, Setare1-2008, Saaidi-2012}.

Moreover, some other models can be referred to as modified teleparallel models which provides an appropriate alternative for describing the accelerated expansion of the universe. Einstein first performed by the mathematical structure of the distant parallelism a unity between electromagnetism and gravity, which was introduced as the teleparallel gravity theory \cite{Einstein_1928}. Electromagnetic and gravity are two sides of the same coin, i.e., gravity and electromagnetism are represented by two mathematical structures, symmetric and antisymmetric, respectively. In this theory, the space-time is characterized by a linear relationship without curvature by a metric tensor field. The tensor of this metric is defined as a dynamical tetrad field. In general relativity, the tetrad fields are a set of diagonal vector fields. One dimension of them is time-like, however, any other three are space-like that are defined on the Lorentz curvature which is a physical concept as a model of space-time. Thus a tetrad field is naturally used to define a linear Weitzenb\"{o}ck connection, which represents a torsion connection without curvature \cite{Weitzenbock_1923, Linder_2010, Myrzakulov_2011, Li_2011, Myrzakulov_2012, Harko_2011, Nojiri_2005}. Also, a tetrad field can naturally be used to define a Riemannian metric and be based on the Levi-Civita connection to show the curvature connection. Due to the universality of gravitational interactions, this geometrical structure may be linked to gravity. Therefore, the teleparallel gravity scenario is replaced rather than general relativity by using the transformation of the tetrad components with the metric components. This means that the curvature term in general relativity is changed to a torsion term in the teleparallel scenario.

Other models of teleparallel gravity are introduced as gravity theories of  $f(T, \mathcal{T})$ model and $f(T, B)$ model, in which $T$, $\mathcal{T}$, and $B$ are the torsion scalar, a trace of the matter energy-momentum tensor and boundary term, respectively \cite{Harko_2014, Rezaei_2017, Bahamonde1_2017, Pourbagher_2019}. These models have less mathematical complexity as well as good agreement with observational data to describe the accelerated expansion of the universe. Between these two models, the gravitational model $ f (T, B) $ is more prominent, due to that one simultaneously recovers both the models of $f(T)$ gravity and $f(R)$ gravity, i.e., provides equivalence between torsion-curvature. Therefore, the $f(T, B)$ gravity can be an alternative for dark energy in both scenarios of teleparallel and general relativity.

In addition, for a more realistic universe, we consider an anisotropic fluid as another category in this regard, i.e., the universe is dominated the $f (T, B)$ scenario with the viscous fluid. \cite{Sadeghi_2013, Pourhassan-2013, SaadatmBB-2013, Amaniali-2013}. This is a motivation to study the dark energy with bulk viscosity on the interactive $f(T, B)$ scenario by correspondence to the new agegraphic. Therefore, in this paper, we intend to study the $f(T, B)$ gravity in the presence of bulk viscosity by using the interacting model between components of the universe. In what follows, we fit the astronomical data with the Hubble parameter derived from the power-law for the scale factor. By using the connection between gravitation and black hole thermodynamics, the validity of the present model using the second law of thermodynamics is investigated \cite{Bardeen-1973, Hawking-1975, Zubair-2017, Bahamonde-2015}.

Therefore, the above material provided an impetus for the study of the cosmic acceleration and its stability by using the scenario of the viscous $f(T, B)$ gravity. Interestingly, this study is performed by correspondence between the new agegraphic and the modified teleparallel gravity in an interactive model, and finally, we validate the present study using the generalized second law of thermodynamics in thermodynamics equilibrium.

This paper is organized as follows:\\
In Sec. \ref{II}, we will study the new agegraphic model. In Sec. \ref{III}, we will explore the $f(T,B)$ gravity in the flat-FRW metric by bulk viscosity fluid, and also obtain the corresponding Friedmann equations. In Sec. \ref{IV}, we consider an interacting model between the components of the universe, and then the Hubble parameter is fitted to astronomical data. In Sec. \ref{V}, we correspond the $f(T, B)$ gravity with the new agegraphic model, as well as the cosmological parameters will obtain and then the stability of the model is investigated. In Sec. \ref{VI} we will examine the validity of the current model by the generalized second law of thermodynamics. Finally, in Sec. \ref{VII} we will present the results and conclusions for the model.

\section{The  new agegraphic dark energy model}\label{II}

In this section, we will review the studies on the agegraphic model. It is well established today that, in general relativity, physical quantities can be measured without any limitations, whereas in quantum mechanics measurements of some physical quantities have limitations. As based on the uncertainty principle the quantum fluctuations of space-time to measure distance $t$ (with the speed of light $c = 1$) in the Minkowski space-time has been done with accuracy
\begin{equation}\label{delt1}
\delta\,t=\lambda\,{t_{{p}}}^{2/3}{t^{1/3}},
\end{equation}
where  $t_{{p}}$  is Planck's reduced time and $\lambda$  is a dimensionless quantity \cite{Karolyhazy-1966, Karolyhazy-1982, Karolyhazy-1986, Maziashvili-2007, Maziashvili1-2007}. Here for simplicity we used $c = \hbar  = {k_d} = 1$. Thus the fluctuation in the energy density of Minkowski space-time metric is $\rho_{{DE}} \approx {\frac {1}{{\kappa}^{2}{t}^{2}}}$ with  $\kappa^2 = 8 \pi G$, in which $G$ is the Newton's gravitational constant. The origin of the agegraphic dark energy is considered based on above energy density, $t$ as the age of the universe and $\tau$  is conformal time can be evaluated by
\begin{equation}\label{tau1}
\tau  = \int {\frac{{dt}}{a} = \int\limits_0^a {\frac{{da}}{H{a^2}}} },
\end{equation}
 where $\dot \tau  = \frac{1}{a}$. Thus the fluctuation in the energy density of metric is
\begin{equation}\label{1}
{\rho _{DE}} = \,\frac{{3{n^2}}}{{{\kappa^2}{\tau ^2}}},
\end{equation}
where the numerical coefficient, $3 n^2$ is used to parameterize some uncertainties, such as quantum field types.

\section{The $f(T,B)$ gravity in the presence of bulk viscosity}\label{III}

As we know, teleparallel gravity theory only has a scalar torsion term, which is expressed in the absence of curvature in general relativity. In the modified teleparallel theory, $T$  converted to $f(T)$ that is comparable with $f(R)$ gravity. However, $f(T)$  and $f(R)$ gravity theories are not equivalent while the $f(T)$ gravity theory is a good alternative for the $f(R)$ gravity theory. In order to compose these theories, following action is introduced,
\begin{equation}\label{2}
S=\int d^4x \, e \, \left(\frac{f(T,B)}{\kappa^2} \, +\mathcal{L}_m\right),
\end{equation}
where $e$ is the determinant of tetrad components, $\mathcal{L}_m$ is the lagrangian  of the matter, $T$  is the scalar torsion and $B = \frac{2}{e}{\partial _\mu }(e{T^\mu })$ is the boundary term, in which ${T^\mu }$ as the torsion tensor that can be define by ${T_\mu } = T_{\nu \mu }^\nu $. By variation of the action \eqref{2} with respect to tetrad field we obtain,
\begin{eqnarray}\label{3}
\begin{aligned}
  2e\delta _\nu ^\lambda {\nabla ^\mu }{\nabla _\mu }{\partial _B}f - 2e{\nabla ^\lambda }{\nabla _\nu }{\partial _B}f + eB{\partial _B}f\delta _\nu ^\lambda  + 4e\left( {{\partial _\mu }{\partial _B}f + {\partial _\mu }{\partial _T}f} \right){S_\nu }^{\mu \lambda } +  \hfill \\
  4e_\nu ^a{\partial _\mu }\left( {e{S_a}^{\mu \lambda }} \right){\partial _T}f - 4e{\partial _T}f{T^\sigma }_{\mu \nu }{S_\sigma }^{\lambda \mu } - ef\delta _\nu ^\lambda  = 16\pi e\mathcal{T}_\nu ^\lambda ,\,\,\,\,\, \hfill \\
  \end{aligned}
\end{eqnarray}
where $\mathcal{T}_\nu ^\lambda  = e_\nu ^a\mathcal{T}_a^\lambda$ is the energy-momentum tensor of the matter. In the teleparallel theory the components of tetrad field ${e_a}({x^\mu })$ are the dynamical variables which form an orthonormal basis for the tangent space at each point ${x^\mu }$ of the space-time manifold. We can write metric in terms of tetrad field as ${g_{\mu \nu }} = {\eta _{AB}}\,e_{\,\,\,\mu }^A\,e_{\,\,\,\nu }^B$ in which ${\eta _{AB}} = diag( - 1,1,1,1)$. Also the relation between the tetra fields determinant and determinant of the  metric  is $e = \det ({e^A}_\mu ) = \sqrt { - g}$. Here, Greek letters  run over from 0 to 3 to denote space-time components while the Latin alphabets run over from 0 to 3 to describe components of tangent space. Now to describe this model, we consider the flat universe by the following Friedmann-Robertson-Walker (FRW) metric,
\begin{equation}\label{4}
d{s^2} =  - d{t^2} + {a^2}(t)\left( {d{x^2} + d{y^2} + d{z^2}} \right),
\end{equation}
where, $a(t)$ is the scale factor. This metric can be constructed by tetra field components as ${e^a}_\mu  = diag(1,a,a,a)$. The torsion scalar and the torsion tensor are respectively defined as
\begin{subequations}\label{56}
\begin{eqnarray}
 &T = {S_\rho }^{\mu \nu }\,{T^\rho }_{\mu \nu }, \label{5}\\
 &{T^\rho }_{\mu \nu } = {e_A}^\rho \,\left( {{\partial _\mu }{e^A}_\nu  - {\partial _\nu }{e^A}_\mu } \right), \label{6}
 \end{eqnarray}
\end{subequations}
where
\begin{equation}\label{srho1}
S_\rho ^{\mu \nu } = \frac{1}{2}\,\left( {K_\rho ^{\mu \nu } + \delta _\rho ^\mu \,T_\alpha ^{\alpha \nu } - \delta _\rho ^\nu \,T_\alpha ^{\alpha \mu }} \right),
\end{equation}
and
\begin{equation}\label{krho1}
K_\rho ^{\mu \nu } =  - \frac{1}{2}\,\left( {{T^{\mu \nu }}_\rho  - T_\rho ^{\nu \mu } - T_\rho ^{\mu \nu }} \right),
\end{equation}
are defined as antisymmetric tensor and contortion tensor, respectively.

The torsion scalar and the boundary term obtain respectively as
\begin{subequations}\label{th1}
\begin{eqnarray}
 &T = 6{H^2},\label{th1-1}\\
 &B = 6\left( {\dot H + 3{H^2}} \right),\label{th1-2}
 \end{eqnarray}
\end{subequations}
where $H = \frac{{\dot a}}{a}$ is Hubble parameter.
It should be noted that the curvature is relate to torsion scalar as, $R = B - T = 6\dot H\, + \,12{H^2}$ \cite{Bahamonde-2017, Bahamonde-2018}.

Now we consider a bulk viscosity in the evolution of universe as a realistic theory. For this purpose, we can write energy–momentum tensor in the presence of the bulk viscosity as
\begin{equation}\label{Tij1}
\mathcal{T}_i^j=(\rho_{tot} + p_{tot} + p_{bulk})u_i u^j - \left(p_{tot} + p_{bulk}\right)\,  \delta_i^j,
\end{equation}
where the total energy density $\rho_{tot}$ and the total pressure $p_{tot}$ are related to fluid inside the universe, and $p_{bulk} = -3 \xi H$ expresses the pressure of bulk viscosity in which $\xi$ is a positive constant for bulk viscosity, as well as the $4$-velocity $u_\mu$ is $u^i$ = (+1,0,0,0) and can be written $ u_i u^j$ = 1. However, elements of energy--momentum tensor be written in the following form
\begin{equation}\label{tau2}
 \mathcal{T}_0^0 = \rho_{tot}, \\
 \mathcal{T}_1^1 = \mathcal{T}_2^2 = \mathcal{T}_3^3 = -p_{tot} + 3 \xi H.
\end{equation}

The above relationships lead to the following Friedmann equations,
\begin{subequations}\label{78}
\begin{eqnarray}
 &- 3{H^2}\left( {3{\partial _B}f + 2{\partial _T}f} \right) + 3H{\partial _B}\dot f - 3\dot H{\partial _B}f + \frac{1}{2}f = {\kappa ^2}{\rho _{tot}},\label{7}\\
 &- \left( {3{H^2} + \dot H} \right)\left( {3{\partial _B}f + 2{\partial _T}f} \right) - 2H{\partial _T}\dot f + {\partial _B}\ddot f + \frac{1}{2}\,\,f =  - {\kappa ^2}\left({p_{tot}}-3 \xi H \right).\label{8}
 \end{eqnarray}
\end{subequations}

We can obtain the total continuity equation from the aforesaid Friedmann equations in the presence of bulk viscosity in the following
\begin{equation}\label{9}
{\dot \rho _{tot}} + 3{\mkern 1mu} H{\mkern 1mu} ({\rho _{tot}} + {\overline{p}_{tot}}) = 0,
\end{equation}
where $\overline{p}_{tot}= {p_{tot}}-3 \xi H$. Now we consider the total energy density and the total pressure of the cosmological fluid of the whole universe as a combination of matter and torsion term as,
\begin{subequations}\label{1011}
\begin{eqnarray}
 &{\rho _{tot}} = {\rho _m} + {\rho _{TB}},\label{10}\\
 &{\overline{p}_{tot}} = {p_m} + {p_{TB}}-3 \xi H,\label{11}
 \end{eqnarray}
\end{subequations}
in here we take $\rho _{TB}$ and $p _{TB}$  as an alternative for dark energy, thus using of Eqs. \eqref{78} and \eqref{1011} we can obtain the energy density and the pressure of the dark energy as below
\begin{subequations}\label{1213}
\begin{eqnarray}
 &{\rho _{TB}} = \frac{1}{{{\kappa ^2}}}\left( { - 3{H^2}\left( {3{\partial _B}f + 2{\partial _T}f} \right) + 3H{\partial _B}\dot f - 3\dot H{\partial _B}f + \frac{1}{2}f} \right) - {\rho _m},\label{12}\\
 &{\overline{p}_{TB}} = \frac{{ - 1}}{{{\kappa ^2}}}\left( { - \left( {3{H^2} + \dot H} \right)\left( {3{\partial _B}f + 2{\partial _T}f} \right) - 2H{\partial _T}\dot f + {\partial _B}\ddot f + \frac{1}{2}f} \right) - {p_m},\label{13}
 \end{eqnarray}
\end{subequations}
where $\overline{p}_{TB}={p}_{TB}-3 \xi H$. In what follows, the equation of state (EoS) for the dark energy is obtained as
\begin{equation}\label{14}
\omega _{TB} = \frac{p_{TB}}{\rho _{TB}}.
\end{equation}
It is worth noting that the EoS is dependent on the $f(T,B)$ model and the viscous fluid.

\section{Interacting model and observational data}\label{IV}
Now, as a realistic theory, we assume that the contents of the universe interact with each other. This means that there is an energy flow between the components of matter and dark energy. So separately, the continuity equations are written from Eq. \eqref{9} for components of matter and dark energy in the following form
\begin{subequations}\label{contin2}
\begin{eqnarray}
 &\dot{\rho}_m+3 H \left(1+\omega_m \right)\rho_m= Q,\label{contin2-1}\\
 &\dot{\rho}_{TB}+3 H \left(\rho_{TB} + \bar{p}_{TB}\right)= -Q,\label{contin2-2}
\end{eqnarray}
\end{subequations}
where $\omega_m=\frac{p_m}{\rho_m}$ is the matter equation of state and $Q$ is interaction term between  the contents of the universe, i.e., when $Q$ to be positive, the energy flow is transferred from dark energy to the matter, and vice versa. In this job, we take the interaction term as $Q = 3 b^2 H \left(\rho_m+\rho_{TB}\right)$ in which $b$ is intensity of the energy flow transfer.

In order to solve the aforesaid equations we consider a specific and precise solution by the power-law for the scale factor in the following form
\begin{equation}\label{at1}
  a(t) = {a_0} {\left(\frac{t}{t_0}\right)^m},
\end{equation}
 where ${a_0}$ is the scale factor of the present universe, $t_0$ is the present age of the universe, and $m$ is dimensionless positive coefficient. The Hubble parameter obtains as
\begin{equation}\label{hubpar1}
  H = \frac{m}{t},
\end{equation}
to insert the present Hubble parameter $H_0 = 68 \pm 2.8 \, km\,s^{-1}\,Mpc^{-1}$ into \eqref{hubpar1} we can obtain the current age of universe as follows:
\begin{equation}\label{hubpar2}
  t_0 = \frac{m}{H_0},
\end{equation}
where in here coefficient $m$ is introduced as correction factor. On the other hand, the relation between the redshift and the scale factor is written by
\begin{equation}\label{hubpar3}
   a(t)=\frac{a_0}{1+z},
\end{equation}
so that we obtain the Eqs. \eqref{at1}-\eqref{hubpar3} the relation between the Hubble parameter and the redshift parameter in the following form
\begin{equation}\label{hubpar4}
   H(z)=H_0  (1+z)^{\frac{1}{m}}.
\end{equation}

\begin{figure}[h]
\begin{center}
\includegraphics[scale=.3]{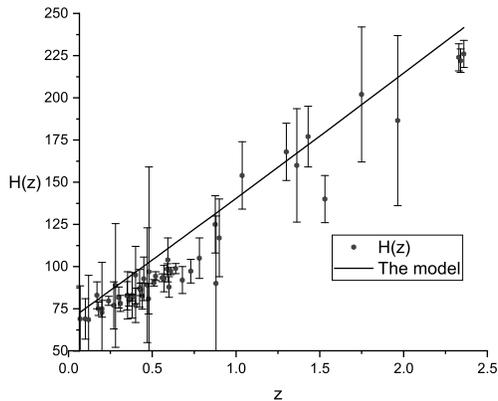}
\caption{The Hubble parameter in terms of the redshift data (diamond $+$ error bar) and the current model (line).}\label{Fig1}
\end{center}
\end{figure}

Now by using the 51 supernova data that they have gathered from Refs. \cite{Farooq_2017, Simon_2005, Stern_2010, Moresco_2012, Blake_2012, Font_2014, Delubac_2015, Moresco_2015, Alam_2016, Moresco_2016, Magana_2018, Pacif_2017}, we fit the equation of Hubble parameter \eqref{hubpar4} by these data. Thus, we obtain the correction factor as $m = 0.956$ by fitting the observational data, which it can be seen in Fig. \ref{Fig1}.

 The interesting point is that by inserting the obtained measurement of correction factor into Eq. \eqref{hubpar2} we can find the current age of universe as $t_0 = 13.75 \, Gyr$. Therefore, the mentioned astronomical date and test of measurement $t_0$ be introduced to the observational constraints in this matter. Then Eq. \eqref{at1} is a suitable choice for the study of the current job.

\section{Correspondence between $f(T,B)$ gravity and new agegraphic}\label{V}
In this section, we aim to correspond to the origin of dark energy that is derived from the $f(T,B)$ gravity to the new agegraphic. By inserting the Eq. \eqref{hubpar1} into \eqref{th1-1} and to calculate the time derivative of Eq. \eqref{hubpar1}, the scalar torsion and the time derivative of Hubble parameter are obtained in terms of time as
\begin{subequations}\label{hubst1}
\begin{eqnarray}
 T =   \frac{6 m^2}{t^2},\label{hubst1-1}\\
 \dot H =  - \frac{m}{t^2}.\label{hubst1-2}
\end{eqnarray}
\end{subequations}
By using the Eq. \eqref{tau1}, the conformal time is evaluated as
\begin{equation}\label{17}
\tau  = \int_0^t {\frac{{dt}}{a}}  = \frac{t_0^m}{a_0 (1-m)}\, t^{1 - m}.
\end{equation}

As it is mentioned, in this job the source of dark energy appear from $f(T, B)$ gravity and new agegraphic. For this purpose, we take that the nature of dark energy is due to fluctuations of space-time  that lead to the calculation of the new agegraphic energy density \eqref{1} in terms of the conformal time. Nevertheless, we consider the correspondence between the energy density of teleparallel gravity, $\rho_{TB}$ and the energy density of new agegraphic, $\rho_{DE}$ as $\rho_{DE} \equiv \rho_{TB}$. In that case, by Eqs. \eqref{1} and \eqref{17} we will have
\begin{equation}\label{18}
 {\rho _{DE}} =\rho _{TB}= \frac{3 \, {n^2}\, H_0^{2 m}\,a_0^2 \, (1 - m)^2}{\kappa ^2 \, m^{2m}}\frac{1}{t^{2(1 - m)}}.
\end{equation}
To use the Eqs. \eqref{contin2-1}, \eqref{hubpar1} and \eqref{18} we can obtain $\rho_m$ as below
\begin{equation}\label{19}
\rho _m = \frac{9 \, n^2 \,H_0^{2 m}\, a_0^2 \, b^2 \,(1 - m)^2}{\kappa ^2 \, \sigma_1 \, m^{2 m-1}}\frac{1}{ t^{2(1 - m)}}+c\, t^{-3 m \sigma_2}
\end{equation}
where $\sigma_1  =  - 3m\,{b^2} + 3m\,{\omega _m} + 5m - 2$, $\sigma_2 = 1+\omega_m-b^2$ and $c$ is  integral constant. By replacing of Eqs. \eqref{hubpar1}, \eqref{th1}, \eqref{18} and \eqref{19} into Eq. \eqref{12} and change of variables ${\partial _B}f = \frac{{\dot f}}{{\dot B}}$ and ${\partial _T}f = \frac{{\dot f}}{{\dot T}}$, we can obtain $f(T(t), B(t))=f(t)$ in the following form
\begin{eqnarray}\label{20}
\begin{aligned}
f( t) = c_1\,t^{m_1 + m_2} + c_2\,t^{m_1 - m_2 } + \frac{6\,{n^2}\,H_0^{2 m}\,{a_0}^2\left( {3{b^2}m + \sigma_1 } \right)\left( {3m-1} \right){{\left( {1 - m} \right)}^2}}{\sigma_1 \left( {7{m^2} - 4 m - 1} \right)\, m^{2m}}\frac{1}{t^{  2(1 - m)}}\\
+\frac{4 \, c\, \kappa^2\, (3 m-1)}{-9 m^2 \sigma_2^2 -27 m^2 \sigma_2 + 6 m \sigma_2 + 6 m - 2} \frac{1}{t^{3 m \sigma_2}},
\end{aligned}
\end{eqnarray}
where $c_1$ and $c_2$ are integral constants, and $m_1=\frac{9}{2}m - 1$ and $m_2= \frac{1}{2}\sqrt {81{\kern 1pt} {m^2} - 12{\kern 1pt} m - 4} $. From Eqs. \eqref{th1} and \eqref{hubst1-2} we find
\begin{equation}\label{t2}
t^2 = \frac{6 m}{ 3 T - B}.
\end{equation}
In order to write the obtained function $f(t)$ in terms of  $T$ and $B$, Eq. \eqref{t2} substitute into Eq. \eqref{20} then we have
\begin{eqnarray}\label{21}
\begin{aligned}
  f(T,B) = c_1\,\left(\frac{6m}{ 3 T-B}\right)^\frac{m_1 + m_2}{2} + c_2\,\left(\frac{6m}{ 3 T-B}\right)^\frac{m_1 - m_2}{2} \\
  + \frac{6\,{n^2}\,H_0^{2 m}\,{a_0}^2\left( {3{b^2}m + \sigma_1 } \right)\left( {3m-1} \right){{\left( {1 - m} \right)}^2}}{\sigma_1 \left( {7{m^2} - 4 m - 1} \right)\, m^{1+m}}\left(\frac{ 3 T-B}{6}\right)^{1-m}  \\
  +\frac{4 \, c\, \kappa^2\, (3 m-1)}{-9 m^2 \sigma_2^2 -27 m^2 \sigma_2 + 6 m \sigma_2 + 6 m - 2}\left(\frac{3 T-B}{6m}\right)^{\frac{3}{2} m \sigma_2}
\end{aligned}
\end{eqnarray}

Here we want to draw the variety of the energy density of dark energy and the pressure of dark energy in terms of redshift parameter. Note that in this approach, the differential relationship is written between the time derivative and derivative with respect to redshift parameter as $\frac{d}{dt}=-H (1+z) \frac{d}{dz}$. In that case, by inserting the Eqs. \eqref{18} and \eqref{19} into \eqref{contin2-2} we have
\begin{subequations}\label{rhide2}
\begin{eqnarray}
&\rho _{TB}= \frac{3 \, {n^2}\, H_0^2\, a_0^2 \, (1 - m)^2}{\kappa ^2 \, m^2}(1+z)^{\frac{2(1-m)}{ m}},\label{rhide2-1}\\
&{p} _{TB}=-\frac {n^2\, H_0^2\, a_0^2\, (1-m)^2\, \left(9\,{b}^{4}{m}^{2}+3\,{b}^{2} m \sigma_1 + 5\,m \sigma_1 - 2\, \sigma_1 \right) }{\kappa^{2}\,\sigma_1\,m^3} \left(1+z\right) ^{\frac{2 (1-m)}{m}}\\ \nonumber
&-\frac{c b^2 H_0^{3 m \sigma_2}}{m^{3 m \sigma_2}}(1+z)^{3 \sigma_2}+3 \xi H_0 (1+z)^{\frac{1}{m}},\label{rhide2-2}
\end{eqnarray}
\end{subequations}

 \begin{figure}[h]
\begin{center}
\includegraphics[scale=.3]{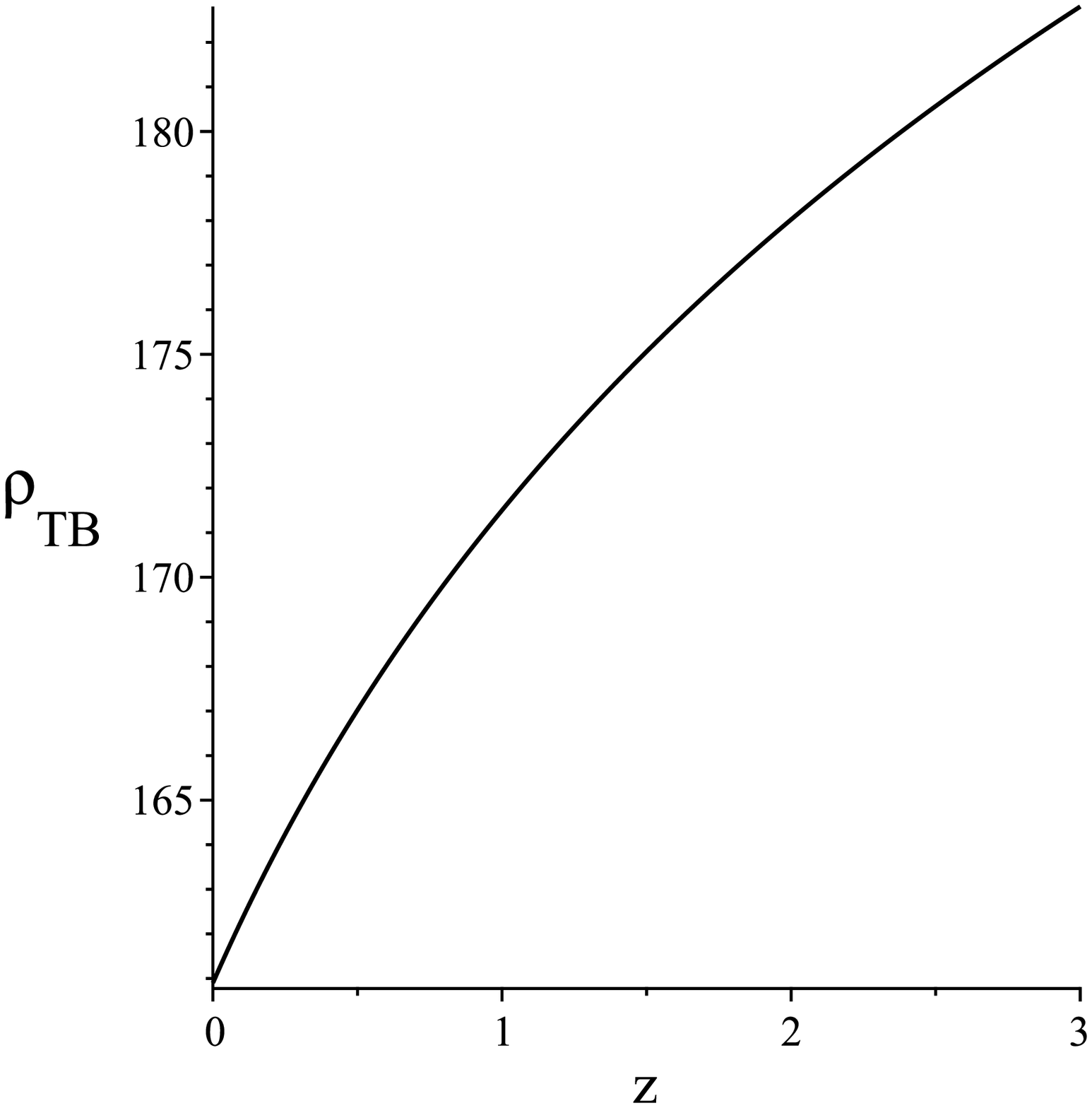} \includegraphics[scale=.3]{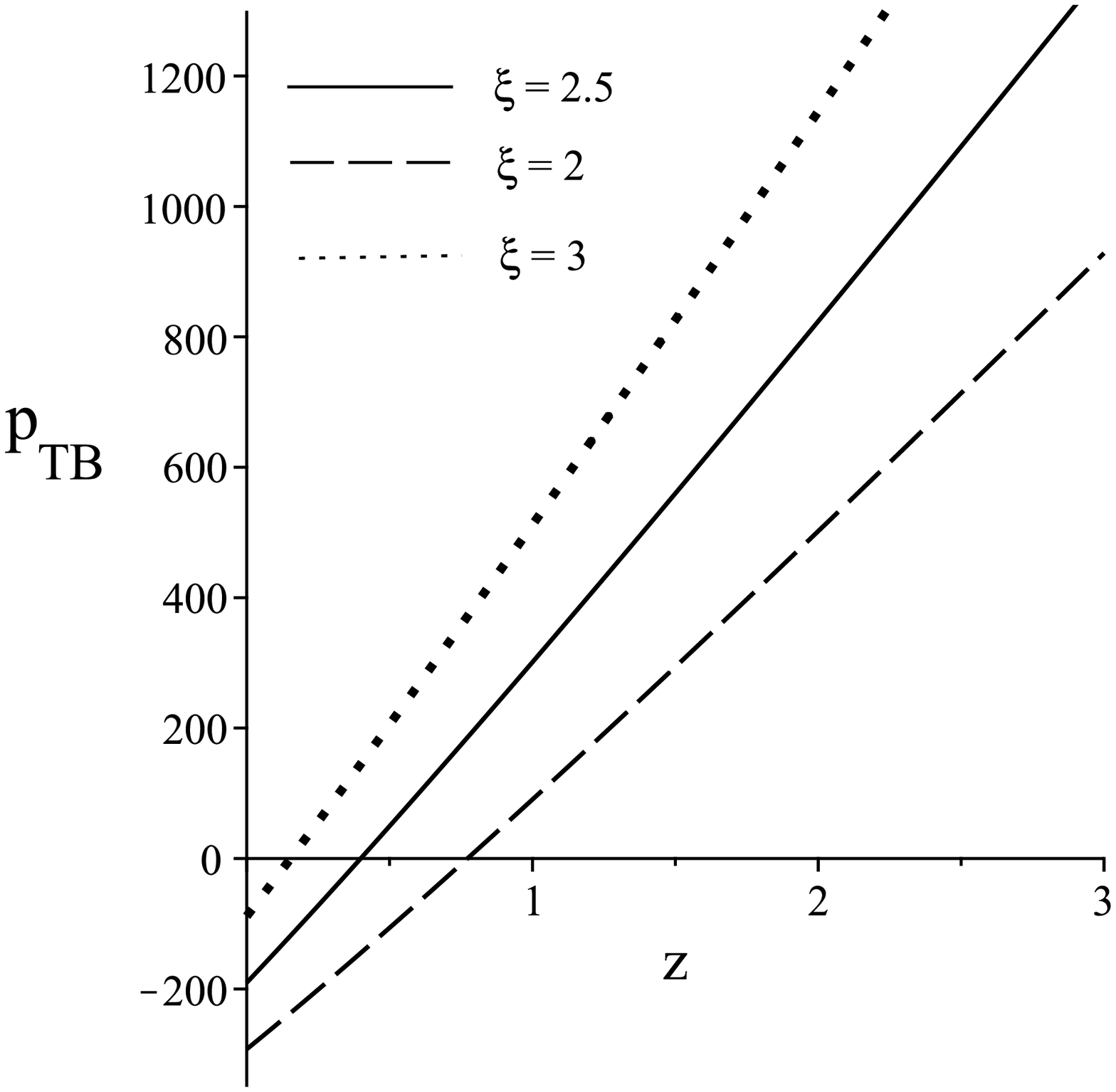}
\caption{The graphs of the energy density ($L^{-1} M T^{-2}$) and the pressure ($L^{-1} M T^{-2}$) of dark energy in terms of redshift parameter by $a_0 = .195$, $b = 0.9$, $m = 0.956$, $n = 3$, $\omega_m = 0.1$, $\xi = 2, 2.5, 3$ and $c = 0.25$.}\label{Fig2}
\end{center}
\end{figure}
\begin{figure}[h]
\begin{center}
\includegraphics[scale=.3]{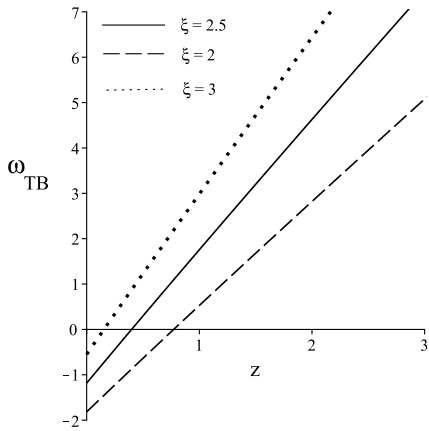}
\caption{The graph of the EoS (dimensional quantity) of dark energy in terms of redshift parameter by $a_0 = .195$, $b = 0.9$, $m = 0.956$, $n = 3$, $\omega_m = 0.1$, $\xi = 2, 2.5, 3$ and $c = 0.25$.}\label{Fig3}
\end{center}
\end{figure}

We draw the graphs of $\rho_{TB}$ and $p_{TB}$ versus to redshift parameter as Fig. \ref{Fig2}. Note that the free parameters play a very important role in drawing graphs. The motivation for our choice is that the universe is undergoing an accelerated expansion phase. For this aim, it has to be the energy density greater than zero and the pressure smaller than zero. Now with these interpretations we choose the free parameters as $a_0 = .195$, $b = 0.9$, $m = 0.956$, $n = 3$, $\omega_m = 0.1$, $\xi = 2, 2.5, 3$ and $c = 0.25$. Here we point out that free parameters $a_0$, $b$, $m$, $\omega_m$ and $c$ are dimensionless and $\xi$ is $pa.s = L^{-1} M T^{-1}$ in SI units or $M^3$ in Planck units.  The Figs. \ref{Fig2} show us that the values of the energy density and the pressure of dark energy are more and less than zero for current universe ($z=0$), respectively. The graph of the energy density is related to free parameters that come from $F(T, B)$ gravity, new agegraphic and interacting model, but one is not dependent on viscosity. On the other hand, the graph of pressure of dark energy is related to $F(T, B)$ gravity, new agegraphic, interacting model and bulk viscosity. The variation of the pressure versus the redshift shows us that the universe from high redshift (early epoch) decrease to a negative value (late epoch). To insert the Eqs. \eqref{rhide2} into \eqref{14}, we obtain the EoS of dark energy (we're not writing here because of the size of his relationship). In what follows, the variation of EoS of dark energy is drawn in terms of the redshift parameter in Fig. \ref{Fig3}. The EoS parameter is a useful parameter in modern cosmology that can justify various epochs of the universe from the Big Bang to the present time. For this purpose, whenever $\omega =1$, $\omega =  \frac{1}{3}$, $0 < \omega < \frac{1}{3}$, $-1 < \omega < -\frac{1}{3}$, $\omega = -1$ and $\omega < -1$ represents stiff fluid, the radiation phase, matter era, accelerated phase or quintessence phase, cosmological constant and phantom era, respectively.  In this case, Fig. \ref{Fig3} shows us that the value of EoS is about $-1.183$ for late time ($z=0$) in $\xi = 2.5$, i. e., we can see that the universe begins from matter epoch, quintessence era, vacuum era and then crossing the phantom separator line for various values of viscosity. Therefore, this issue indicates that the universe is undergoing accelerated expansion as well as one confirms the results obtained in Ref. \cite{Amanullah_2010, Scolnic_2018}.

\begin{figure}[t]
\begin{center}
\includegraphics[scale=.3]{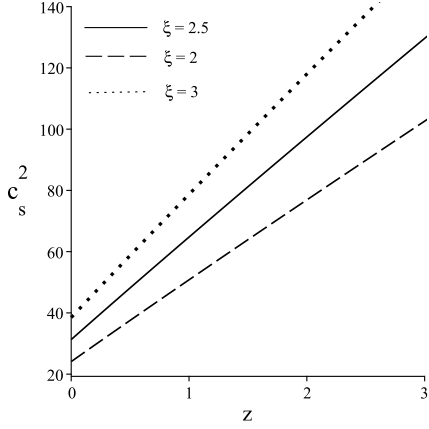}
\caption{The graph of the sound speed (dimensional quantity) in terms of redshift parameter by $a_0 = .195$, $b = 0.9$, $m = 0.956$, $n = 3$, $\omega_m = 0.1$, $\xi = 2, 2.5, 3$ and $c = 0.25$.}\label{Fig4}
\end{center}
\end{figure}

Now we intend to explore the stability of our model, so we need to introduce a useful function named parameter of the squared sound speed, $c_s^2$ as given below
\begin{equation}\label{cs21}
  c_s^2=\frac{\partial p_{TB}}{\partial \rho_{TB}} = \frac{\partial_z p_{TB}}{\partial_z \rho_{TB}},
\end{equation}
where index $z$ is derivative with respect to redshift parameter. The stability condition of the current model is that the sound speed is bigger than zero, i.e., $c_s^2 > 0$, otherwise, it has an unstable condition. In that case, we draw the variation of sound speed parameter versus redshift parameter in Fig. \ref{Fig4}. Fig. \ref{Fig4} shows us that the value of the sound speed parameter decreases from the early era and goes towards the late epoch by a positive value in various values of viscosity. Therefore, there is the stability condition from the early phase to the late time, because the sound speed is bigger than zero in all eras.

\section{Thermodynamics analysis}\label{VI}
In this section the second law of thermodynamics is investigated for the modified teleparallel gravity model. Here we assume that the boundary of the universe is bounded by the radius of the apparent horizon \cite{Bekenstein_1973, Hawking_1975}. By using of the power-law model of the scale factor \eqref{at1}, we can obtain the radius of apparent horizon in Flat-FRW metric by
\begin{equation}\label{22}
\bar r = \frac{1}{H} = \frac{t}{m},
\end{equation}
and then its derivative with respect to time is given below
\begin{equation}\label{rdot1}
\dot {\bar r} =  - \frac{{\dot H}}{H^2} = \frac{1}{m}.
\end{equation}

According to the second law of thermodynamics, the total entropy of the universe which is consists of entropies of matter inside and on the boundary of the apparent horizon must always increase in terms of time evolution. Therefore, the Hawking temperature on the boundary of horizon is calculated as follows:
\begin{equation}\label{23}
{T_h} = \frac{1}{{2\pi \bar r}}\left( {1 - \frac{{\dot{ \bar r}}}{{2H\,\bar r}}} \right) = \frac{2 m - 1}{4 \pi t},
\end{equation}
since the Hawking temperature is always positive, then we must have the correction factor $m > 0.5$. Of course, the correction factor obtained by observational constraints as $m = 0.956$ which ones bigger than obtained condition. The Gibbs equation is written for the entropy of the contents inside apparent horizon in the following form
\begin{equation}\label{24}
{T_h}d{S_{in-hor}} = d({\rho _{tot}}\,V) + {\overline{p}_{tot}}\,dV = V\,d{\rho _{tot}} + ({\rho _{tot}} + {\overline{p}_{tot}})dV,
\end{equation}
where ${\rho _{tot}}$ and ${\overline{p}_{tot}}$ come from the equations \eqref{1011}, and $V = \frac{4}{3}\pi {\bar r^3}$. To differentiate of the Eq. \eqref{24} with respect to time and using Eqs. \eqref{9} and \eqref{10} we will have
\begin{equation}\label{26}
{T_h}{\dot S_{in-hor}} = \frac{4 \pi \bar r^2}{3 H}(\dot{\rho} _{TB} + \dot{\rho} _{m})(1-\dot {\bar r}),
\end{equation}
and using the Eqs. \eqref{18} and \eqref{19} will be
\begin{equation}\label{tsinhor2}
{T_h}{\dot S_{in-hor}} = \frac{3 (1-m)}{m^4}\left(\frac{n^2 H_0^{2 m} a_0^2 (1-m)^3(3 b^2 m+\sigma_1)}{G \,\sigma_1\, m^{2 m}}t^{2 m} + 4 \pi c \,m \,\sigma_2\, t^{2-3 m \sigma_2}\right).
\end{equation}

On the other, Bekenstein–Hawking horizon entropy is written in general relativity as ${S_{on-hor}} = \frac{A}{4 G}$ in which $A = 4\pi {\bar r^2}$ is introduced as the area of the apparent horizon. In that case, we have
\begin{equation}\label{tenton1}
 T_h \dot{S}_{on-hor} = \frac{\dot{\bar{r}}}{G}\left(1-\frac{\dot{\bar{r}}}{2}\right)=\frac{2 m-1}{2 G m^2}.
\end{equation}

As mentioned above, the total entropy should not decrease in terms of time evolution. Therefore, the validity of the generalized second law of thermodynamics in thermodynamics equilibrium (temperature of the universe inside and on the boundary of the apparent horizon is the same) corresponds to the following condition
\begin{equation}\label{antropytot}
 \dot{S}_{tot} = \dot{S}_{in-hor}+\dot{S}_{on-hor}\geq 0,
\end{equation}
it means that we were able to obtain a constraint on the choice of free parameters values during of time evolution. In that case, we can write this constraint for the free parameters in the following form
\begin{equation}\label{constraint2}
 T_h \dot{S}_{tot} =\frac{3 (1-m)}{m^4}\left(\frac{n^2 H_0^{2 m} a_0^2 (1-m)^3(3 b^2 m+\sigma_1)}{G \,\sigma_1\, m^{2 m}}t^{2 m} + 4 \pi c \,m \,\sigma_2\, t^{2-3 m \sigma_2}\right) + \frac{2 m-1}{2 G m^2} > 0.
\end{equation}
 \begin{figure}[t]
\begin{center}
\includegraphics[scale=.3]{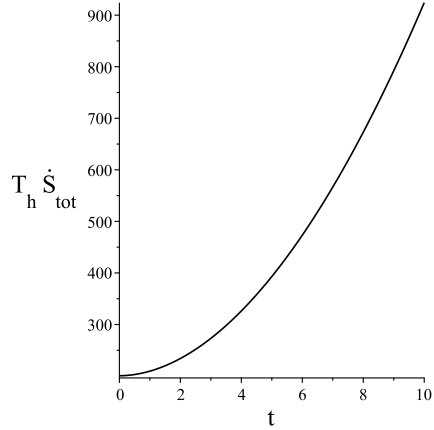}
\caption{The graph of the total entropy in terms of time evolution by $a_0 = .195$, $b = 0.9$, $m = 0.956$, $n = 3$, $\omega_m = 0.1$ and $c = 0.25$.}\label{Fig5}
\end{center}
\end{figure}
It should be noted that by inserting the free parameter values (see the previous section) into the above constraint, the variation of the total entropy versus time evolution is positive as it is shown in Fig. \ref{Fig5}. Therefore, we conclude from Fig. \ref{Fig5} that the total entropy of the universe during the time evolution is increasing.

\section{Conclusion}\label{VII}
In this paper, we studied the new agegraphic model as an alternative to the dark energy and obtained the energy density in terms of conformal time. On the other hand, the $f(T, B)$ gravity model has been explored with a viscous fluid in the flat-FRW universe as a source of dark energy. The most important advantage of the $f(T, B)$ model is that boundary term $B$ relates the Ricci scalar with the torsion scalar as $B=R+T$.  This means that the $f(T, B)$ model alone covers only one of the gravity models of  $f(T)$ and the $f(R)$ based on the Weitzenb\"{o}ck connection and the Levi-Civita connection, respectively. Next, we obtained the Friedmann equations in the presence of viscous fluid, then we considered the contents of the universe by components of matter and dark energy. Immediately afterward, the continuity equations in terms of universe components have separately written by applying an interactive term as $Q= 3 b^2 H (\rho_m + \rho_{TB})$.

In order to solve the job, we take the power-law for the scale factor, then the corresponding Hubble parameter fitted with 51 supernova data, and thence  the age of universe found as $t_0 = 13.75 \,Gyr$. In what follows, we constructed the correspondence between the energy density of teleparallel
gravity and the energy density of new agegraphic, and then we obtained function $f(T, B)$ as functional of $T$ and $B$. Afterwards we have written the energy density and the pressure of dark energy in terms of the redshift parameter, and we have drawn the cosmological parameters such as the energy density, the pressure and the EoS of dark energy versus redshift parameter. Interestingly, the free parameters were chosen on the basis that the universe is
undergoing an accelerated expansion phase. So, the measurement of the EoS parameter is equal to $-1.183$ in late time evolution for $\xi = 2.5$ and is compatible with observational data. For a more complete discussion, the stability of the current model has been analyzed by the sound speed parameter and saw that there is a stability condition in late time. Finally, the generalized second law of thermodynamics explored in a flat-FRW universe by the apparent horizon.  Therefore, we concluded that the variation of the total entropy is increasing versus time evolution in the thermodynamic equilibrium condition for selected free parameters.


\end{document}